\documentclass{article}

\usepackage[english]{babel}

\usepackage[letterpaper,top=2cm,bottom=2cm,left=3cm,right=3cm,marginparwidth=1.75cm]{geometry}

\usepackage{amsmath}
\usepackage{graphicx}
\usepackage[colorlinks=true, allcolors=blue]{hyperref}
\usepackage{amsmath}
\usepackage{amsfonts}
\usepackage{booktabs} % For professional looking tables
\usepackage{longtable} % For tables that may span multiple pages

\title{Large language models can help boost food production,\\ but be mindful of their risks}

\author{\small \bf Djavan De Clercq\\
\textit{\small University of Oxford}\\
\and
\small\bf  Elias Nehring\\
\textit{\small ETH Zurich}\\
\and
\small\bf  Harry Mayne\\
\textit{\small University of Oxford}\\
\and
\small\bf  Adam Mahdi\\
\textit{\small University of Oxford}
}

\date{}

\begin{document}
\maketitle

%%%%%%%%% ABSTRACT 
\begin{abstract}

Coverage of ChatGPT-style large language models (LLMs) in the media has focused on their eye-catching achievements, including solving advanced mathematical problems and reaching expert proficiency in medical examinations. But the gradual adoption of LLMs in agriculture, an industry which touches every human life, has received much less public scrutiny. In this short perspective, we examine risks and opportunities related to more widespread adoption of language models in food production systems. While LLMs can potentially enhance agricultural efficiency, drive innovation, and inform better policies, challenges like agricultural misinformation, collection of vast amounts of farmer data, and threats to agricultural jobs are important concerns. The rapid evolution of the LLM landscape underscores the need for agricultural policymakers to think carefully about frameworks and guidelines
that ensure the responsible use of LLMs in food production before these technologies become so ingrained that policy intervention becomes challenging.

\end{abstract}

\tableofcontents

%%%%%%%%% INTRODUCTION 
\section{Introduction}

In 2023, generative AI technologies such as large language models (LLMs) took the world by storm due to the popularity of tools such as ChatGPT. Like other forms of artificial intelligence, the capability of generative AI relies on training with vast amounts of data. It is capable of generating new content - text, images, audio, videos, and even computer code - instead of simply categorizing or identifying data like other AI systems \cite{foster_generative_2023}. Bold projections about the potential of generative AI such as those from institutions like Sequoia Capital, which in late 2022 claimed that ``every industry that requires humans to create original work is up for reinvention”, helped AI startups attract one out of every three dollars of venture capital investment in the U.S. the following year \cite{gpt-3_generative_2022, Hu2024Jan}. 

While media headlines have focused on eye-catching examples of the impact of generative AI -- such as Google’s Med-Palm 2 achieving clinical expert -- level performance on medical exams, Microsoft’s BioGPT reaching human parity in biomedical text generation, and Unity Software's enhancements of video game realism - emerging applications of LLMs in the field of agriculture have received much less public attention \cite{Tong2023Mar, Singhal2023May, Luo2022Oct}. Examples which have mostly gone unnoticed by major news outlets include research from Microsoft which demonstrated that LLMs could outperform humans on agronomy examinations, and the gradual roll-out of systems like KissanGPT, a ``digital agronomy assistant" which allows farmers with poor literacy skills in India to interact with LLMs verbally to receive tailored farming advice \cite{Silva2023Oct, BibEntry2024Feb}. These early developments may be precursors to larger-scale applications that could have considerable repercussions for an industry which touches every human life: food production. 

In this short perspective, we present a brief overview of the opportunities and risks associated with increasing the use of large language models across food systems. Given that this is a nascent field, this perspective builds on insight from diverse sources, including journalistic reports, academic preprints, peer-reviewed literature, innovations in startup ecosystems, and lessons from the outcomes of early LLM adoption in fields outside of agriculture. 

%%%%%%%%% OPPORTUNITIES
\section{Opportunities to boost food production with LLMs}

If used judiciously, LLMs may have a role to play in addressing global food production challenges. There will be 10 billion people on Earth by 2050 – around 2 billion more people than there are today – and this may test the global agricultural system and the farmers who comprise it \cite{searchinger_creating_2019}.  Research has warned that agricultural production worldwide will face challenges in meeting global demand for food and fiber, with food demand estimated to increase by more than 70\% by 2050 \cite{ray_yield_2013, valin_future_2014, sishodia_applications_2020}. Such challenges include climate-related pressures and extreme weather, biotic threats such as pests and diseases, decreasing marginal productivity gains, soil degradation, water shortages, nutrient deficiencies, urban population growth, rising incomes, and changing dietary preferences \cite{bailey-serres_genetic_2019, fao_future_2018, national_academies_of_sciences_science_2019}. Here, we outline opportunities for stakeholders across the food production ecosystem to use LLMs to address some of these challenges. 

\subsection{Boosting agricultural productivity}

\paragraph{On-demand agronomic expertise} Language models like GPT-4 could transform the way farmers seek and receive advice in agricultural settings. These models are capable of providing context-specific guidance to users across various fields and convincingly portray specific roles as agricultural experts. They achieve this by inquiring about symptoms, conducting thorough questioning based on responses, and providing useful support \cite{Lai2023Jul, Chen2023May}. LLMs have already achieved high scores on exams for renewing agronomist certifications, answering 93\% of the exam questions correctly in the United States' Certified Crop Advisor (CCA) certification \cite{Silva2023Oct}. 

This ability to simulate expert roles has led to commercial chatbot-based products that emulate agronomists or crop scientists, providing farmers with on-demand advice. Early examples include {\it KissanGPT} \cite{BibEntry2024Feb}, a chatbot which assists Indian farmers with agricultural queries such as “how much fertilizer should I apply to my fields?”; {\it Norm} \cite{Marston2023May}, a chatbot which provides guidance on topics such as pest mitigation strategies and livestock health; Bayer's LLM-powered agronomy advisor, which answers questions related to farm management and Bayer agricultural products \cite{Bayer2024Mar}; and {\it FarmOn} \cite{PodcastRegen}, a startup which aims to provide an LLM-augmented hotline that provides agronomic advice on areas such as regenerative farming. Farmers can interact with some of these digital advisors via web-based platforms, or mobile messaging tools like {\it WhatsApp}, in a number of supported languages. When these chatbots have access to all of the world’s agronomic texts in various languages—including farming manuals, agronomy textbooks, and scientific papers, ``digital agronomists" could place the entirety of the world's agronomic data at farmers' fingertips.

There is emerging evidence that farmers may be willing to engage with such digital tools for farming advice in both developed and developing agricultural regions. One study found that farmers in Hungary and the UK are increasingly leveraging digital information sources -- such as social media, farming forums, and online scientific journals -- over traditional expert advice on topics such as soil management \cite{Rust2022Jan}. In Nigeria, researchers demonstrated that ChatGPT-generated responses to farmers' questions on irrigated lowland rice cultivation were rated significantly higher than those from agricultural extension agents \cite{Ibrahim2024Feb}. One study involving 300 farmers in India's Karnataka state found that providing farmers with access to agricultural hotlines with rapid, unambiguous information by agricultural experts over the phone, tailored to time- and crop-specific shocks, was associated with greater agricultural productivity through the adoption of cost-effective and improved farming practices \cite{Subramanian2021Jun}.

While these examples offer evidence that LLM-based tools can impact agricultural productivity, there may be additional opportunities to improve them by connecting them with real-time information from sources such as satellite imagery or current market data. Such integration could enable them to answer not just straightforward questions like ``how much fungicide should I apply to my corn crop," but also more context-specific queries such as ``how much fungicide should I apply to my crop, considering the rainfall and wind conditions that have affected my fields in the past two weeks and my expected income at current market prices?" An early example in this vein is \textit{LLM-Geo} \cite{Li2023May}, a system which allows the user to interact with earth observation data, such as satellite imagery and climate data, in a conversational manner. Such advancements may be particularly valuable for farmers, as collecting and analyzing satellite and climate data has typically been a task requiring specialized skills in software or programming \cite{Herrick2023Jan}. Moreover, agronomic advice that is  time-sensitive and location-specific tends to be viewed more favourably by farmers than more general responses about agronomy \cite{Ibrahim2024Feb, Kassem2020Oct}. 

In addition to providing information that is both location-specific and time-sensitive, ``digital agronomists" powered by LLMs could also be enhanced with the ability to reason across multiple modes of data, such as images, video, and audio. A farmer could, for instance, upload images of crops affected by pests and engage in verbal dialogue with an LLM to explore potential solutions. For example, one study demonstrated the ability of such a system to understand and engage in discussions about images of pests and diseases affecting Chinese forests. This system showcased the ability to conduct dialogue and address open-ended instructions to gather information on pests and diseases in forestry based on multiple types of data \cite{Zhang2023Oct}.

% In addition to time and location specificity, one study on 382 farmers in Egypt's Dakhalia governorate demonstrates that, to reach their full potential, mobile-based advisory services should be tailored to the socio-economic and educational characteristics of their users, given that farmers' responses to digital information can depend on these factors \cite{Kassem2020Oct}. 

\paragraph{Agricultural extension services} While agronomic knowledge can be directly imparted to farmers via services like KissanGPT, some studies have also highlighted the potential of LLMs in enhancing existing agricultural extension services. Agricultural extension services, which serve as a bridge between research and farming practices, typically involve government or private agencies providing training and resources to improve agricultural productivity and sustainability. Across different countries, these services are administered through a mix of on-the-ground workshops, personalized consulting, and digital platforms, adapting to local needs and technological advancements. Tzachor et al. \cite{Tzachor2023Nov} highlight the limitations of traditional agricultural extension services, such as their limited reach, language barriers, and the lack of personalized information. They propose that LLMs can overcome some of these challenges by simplifying scientific knowledge into understandable language, offering personalized, location-specific, and data-driven recommendations.

\paragraph{Conversational interfaces for on-farm agricultural robotics} Roboticists have highlighted the potential of LLMs to impact human-robot interaction by providing robots with advanced conversational skills and versatility in handling diverse, open-ended user requests across various tasks and domains \cite{Kim2024Jan, Wang2024Jan}. OpenAI has already teamed up with Figure, a robotics company, to build humanoid robots that can accomplish tasks such as making coffee or manipulating objects in response to verbal commands \cite{Figure2024Mar}. For food production specifically, Lu et al.  \cite{Lu2023Apr} argue that LLMs may play a role in helping farmers control agricultural machinery on their farms, such as drones and tractors. LLMs might act as user-friendly interfaces that translate human language commands into machine-understandable directives \cite{Vemprala2023Feb}. Farmers could, for instance, provide intuitive and context-specific instructions to their agricultural robots via dialogue, reducing the need for technical expertise to operate machinery. LLMs could then guide robots to implement optimal farming practices tailored to the specific conditions of the farm, such as directing drones to distribute precise amounts of pesticides or fertilizers, or guiding tractors in optimized plowing patterns, thereby improving crop yields and sustainability. In addition, farmers may be able to leverage the multimodal capabilities  (i.e., processing and integrating different types of information like text, images, and numerical data) of LLMs to engage in dialogue about data collected from agricultural machinery, such as imagery from drones and soil information from harvesting equipment. 

While many studies have placed emphasis on guiding robots on what actions to perform in real-world planning tasks, Yang et al.  \cite{Yang2023Sep} highlight the equal importance of instructing robots on what actions to avoid. This involves clearly communicating forbidden actions, evaluating the robot's understanding of these limitations, and crucially, ensuring adherence to these safety guidelines. This may be critical to ensure that LLM-powered agricultural robots can make informed decisions, avoiding actions that could lead to unsafe conditions for crops and human workers.

\subsection{Accelerating agricultural innovation}

Throughout history, advancement in agriculture has been vital for economic development by boosting farm productivity, improving farmers' incomes, and making food increasingly plentiful and affordable  \cite{Alston2021Jan}.  Agricultural innovation has many features in common with innovation more broadly, but it also has some important differences. First, unlike innovation in manufacturing or transportation, agricultural technology must consider site-specificity due to its biological basis, which varies with climate, soil, and other environmental factors, making technology and innovation returns highly localized. Secondly, industrial technologies are not as susceptible to the types of climatic, bacterial, or virological shocks that typify agriculture and drive obsolence of agricultural technology \cite{Pardey2010Jan}. Large language models may present opportunities to address the specific needs of agricultural innovation with these differences in mind.

\paragraph{Agricultural research and development} In the context of a field where innovation involves software, hardware, biology, and chemistry, there are several avenues through which agricultural innovation may be accelerated. 

%\smallskip
%\noindent
%\textit{Faster development of reproducible agricultural science and software}. 
“Coding assistants" powered by LLMs are impacting the agricultural software development landscape by enabling software developers to produce more quality code in less time. {\it GitHub Copilot}—Microsoft’s coding assistant—is a prominent example of these tools. One study revealed that developers using Copilot completed their coding tasks over 50\% faster than their counterparts who did not have access to the tool \cite{Peng2023Feb}. Another example is {\it Devin}, a software development assistant which can finish entire software projects on its own \cite{Vance2024Mar}. Such tools may accelerate the pace of innovation not only for employees in the private sector, but also for researchers in the agricultural sciences who do not possess a background in computer science or software design \cite{Chen2024Feb}. LLM-enabled coding assistants can help produce well-structured and documented code which supports code reuse and broadens participation in the sciences from under-represented groups. Even if the LLM does not generate fully functional code, tools that suggest relevant APIs, syntax, statistical tests, and support experimental design and analysis may help improve workflows, help manage legacy code systems, prevent misuse of statistics, and ultimately accelerate reproducible agricultural science \cite{morris2023scientists}. 

%\smallskip
%\noindent
%\textit{Faster data processing of vast agriculture-related datasets}. 
Petabytes of data relevant to agrifood research are generated every year, such as satellite data, climate data, and experimental data \cite{Mohney2020Apr, Hawkes2020Mar} and this sheer volume can make data wrangling work an important bottleneck to producing agricultural research \cite{Xu2020Feb}. The ability to issue data queries in natural language to an AI model that would allow it to perform some or all of the steps involved in dataset preparation (or at least to generate code that would handle data processing steps) may provide a boost to the speed of agricultural research that depends on vast earth observation datasets. Research teams from NASA and IBM have started to explore the use of generative AI to tackle earth observation challenges by building \textit{Privthi}, an open-source geospatial foundation model for extracting information from satellite images such as the extent of flooding area \cite{Li2023Sep}. These models have also shown promise in location recognition, land cover classification, object detection, and change detection \cite{Zhang2024Jan}.  

\paragraph{Language models for knowledge synthesis and trend identification} Language models can help synthesize the state of the art in agriculture-related topics to accelerate learning about new areas \cite{Zheng2023Oct}. A search on ScienceDirect for papers related to “climate change effects on agriculture” since 2015, for instance, returns more than 150,000 results. Tools like ChatGPT (and its various plugins such as ScholarAI) and Elicit have demonstrated a strong capacity for producing succinct summaries of research papers. Some of these LLM-based tools are capable of producing research summaries that reach human parity and exhibit flexibility in adjusting to various summary styles, underscoring their potential in enhancing accessibility to complex scientific knowledge \cite{Fonseca2024Jan}.   

In addition to knowledge synthesis, language models may also help find patterns in a field over time and reveal agricultural research trends that may not be apparent to newcomers, or even more established scientists. This could help to pinpoint gaps in the literature to signify areas for future research. For instance, emerging LLM-powered tools such as Litmaps and ResearchRabbit are able to generate visual maps of the relationships between papers, which can help provide a comprehensive overview of the agricultural science landscape and reveal subtle, cross-disciplinary linkages that might be missed when
agricultural researchers concentrate exclusively on their own specialized fields \cite{sulisworo2023exploring}.

Lastly, LLMs can help aspiring researchers to engage in self-assessment through reflective Q\&A sessions, allowing them to test their understanding of scientific concepts \cite{Dan2023Aug,Koyuturk2023Jun}. This is especially valuable for interdisciplinary research that combines agronomic science with other fields such as machine learning, which can be laden with jargon.

\paragraph{Accelerated hypothesis generation and experimentation} Language models like GPT-4 have been shown to possess abilities in mimicking various aspects of human cognitive processes, including abductive reasoning, which is the process of inferring the most plausible explanation from a given set of observations \cite{Glickman2024Jan}. Unlike statistical inference, where a set of possible explanations is typically listed, abductive reasoning typically involves some level of creativity to come up with potential explanations which are not spelled out a priori, much like a doctor who abductively reasons the likeliest explanations of a patient's symptoms before proposing treatment.    

This capability is particularly relevant in fields like agricultural and food system research, where synthesizing information from diverse sources such as satellite, climate, and soil information are key to generating innovative solutions. For instance, consider a scenario in agriculture where a researcher observes a sudden decline in crop yield despite no significant changes in farming practices. An LLM could assist in abductive reasoning by generating hypotheses based on the symptoms described by the researcher, suggesting potential causes such as soil nutrient depletion, pest infestation, or unexpected environmental stressors, even if these were not initially considered by the researcher. 

Fast hypothesis generation can help accelerate the shift towards subsequent experimentation, which can also be aided by LLMs. One study demonstrated an ‘Intelligent Agent system’ that combines multiple LLMs for autonomous design, planning, and execution of scientific experiments such as catalyzed cross-coupling reactions \cite{Boiko2023Apr}. In the context of agriculture, if the goal is to develop a new pesticide, for instance, such a system could be given a prompt like ``develop a pesticide that targets aphids but is safe for bees." The LLM would then search the internet and scientific databases for relevant information, plan the necessary experiments, and execute them either virtually or in a real-world lab setting. This might significantly speed up the research and development process, making it easier to innovate and respond to emerging challenges in agriculture. 

\paragraph{Bridging linguistic barriers in research} English is currently used almost exclusively as the language of science. But given that less than 15\% of the world’s population speaks English, a considerable number of aspiring scientists and engineers worldwide may face challenges in accessing, writing, and publishing scientific research \cite{Drubin2012Apr}. LLMs can help address these disparities by making science more accessible to students globally, regardless of their native language. In addition, LLMs may help non-native English speakers communicate their scientific results more clearly by suggesting appropriate phrases and terminology to ensure readability for their intended audience \cite{Abd-Alrazaq2023Jun}, which could facilitate the inclusion of broader perspectives in scientific research on food systems.

\paragraph{Leveraging language models for feedback} Scientists and engineers may be able to leverage LLMs to receive feedback on the quality of their innovations. One study introduced ``Predictive Patentomics", which refers to LLMs that can estimate the likelihood of a patent application being granted based on its scientific merit. Such systems could provide scientists and engineers with valuable feedback on whether their innovations contain enough novelty to qualify for patent rights, or if they need to revisit and refine their ideas before submitting a new patent application \cite{Yang2023Jun}. Other studies have shown similar potential for LLMs to provide feedback on early drafts of scientific papers \cite{Liu2023Jun, Liang2023Oct, Robertson2023Jun}.

\subsection{Improving agricultural policy}

\paragraph{``Second opinions" on the potential outcomes of agricultural policies}Using software and models that mimic realistic human behavior can provide public servants with clearer insight into the potential repercussions of their policy choices \cite{Steinbacher2021Jul}. For instance, one study used LLMs to incorporate human behavior in simulations of global epidemics, finding that LLM-based agents demonstrated behavioral patterns similar to those observed in recent pandemics \cite{Williams2023Jul}. Another study introduced “generative agents”, or computational software agents that simulate believable human behavior, such as cooking breakfast, heading to work, initiating conversations, and “remembering and reflecting on days past as they plan the next day.” \cite{Park2023Apr} Simulations using LLMs have also exhibited realistic human behavior in strategic games \cite{Guo2023May}. 

Such examples offer early evidence that LLMs may enable public servants to ``test" agricultural policy changes by simulating farmer behavior, market reactions, and supply chain dynamics, allowing them to refine their ideas before putting them into practice. Some governments around the world have already started to investigate the incorporation of generative AI in public sector functions. Singapore launched a pilot program where 4,000 Singaporean civil servants tested an OpenAI-powered government chatbot as an “intellectual sparring partner", assisting in functions like writing and research \cite{Min2023Jul}. Meanwhile, the UK government's Central Digital and Data Office has provided guidelines encouraging civil servants to “use emerging technologies that could improve the productivity of government, while complying with all data protection and security protocols.” \cite{GenAI_UK} 

%\paragraph{Better access to and engagement with government services by citizens involved in agricultural production} 
\paragraph{Enhanced engagement with government services} Some governments have started to leverage LLMs in government-approved chatbots that allow citizens to query government databases and report municipal issues via WhatsApp \cite{OneService, SingGenAI}. Applied to agriculture, such chatbots could enable farmers to ask questions related to subsidy eligibility, legal compliance, or applications for licenses to produce organic food. 

Moreover, government agencies might use chatbots as an interface for receiving feedback from citizens about agricultural policies, programs, or services. This can be done in a conversational manner, making it easier for individuals to voice their opinions or concerns and for public servants to better understand farmers' pain points based on millions of queries  \cite{Small2023Jun}. For example, an agricultural economist at a government agency might use LLMs to assist with engaging in dialogue with a group of farmers, rather than just asking participants to fill out surveys. This may yield socio-agronomic data that is more reflective of real-world scenarios \cite{Jansen2023Sep}.

%\paragraph{Better monitoring of and response to shocks to food production systems}
\paragraph{Monitoring agricultural shocks}
Language models may prove valuable for governments seeking to monitor and respond to shocks in agricultural production and food supply. LLMs are proficient in parsing through vast quantities of news text data and distilling them into succinct, relevant summaries that are on par with human-written summaries \cite{Zhang2023Jan}.

This human-level performance could help warn about global food production disturbances to issue early warnings to populations at risk or guide humanitarian assistance to affected regions. One study analyzed 11 million news articles focused on food-insecure countries, and showed that news text alone could significantly improve predictions of food insecurity up to 12 months in advance compared to baseline models that do not incorporate such text information \cite{Balashankar2023Mar}. Another study built a real-time news summary system called \textit{SmartBook}, which digests large volumes of news data to generate structured ``situation reports", aiding in understanding the implications of emerging events \cite{reddy_smartbook_2023}. If used to monitor shocks to food supply due to droughts or floods, such systems could provide intelligent reports to expert analysts, with timelines organized by major events, and strategic suggestions to ensure effective response for the affected farming populations. 

%%%%%%%%% RISKS 
\section{Potential risks for food systems as LLM use spreads}

Despite the excitement around the potential uses of generative AI across various industries, some offer a more sober assessment of its impact, suggesting that the world's wealthiest tech companies are seizing the sum total of human knowledge in digital form, and using it to train their AI models for profit, often without the consent of those who created the original content. And by the time the implications of these technologies are fully understood, they may have become so ubiquitous that courts and policy-makers could feel powerless to intervene \cite{Klein2023Aug}. More general concerns have been raised about generative AI, such as errors in bot-generated content, fictitious legal citations, effects on employment due to the potential automation of worker tasks, and reinforcement of societal inequalities \cite{Merken2023Jun, Woo2024Feb, noauthor_generative_2023, Eloundou2023Mar, okerlund2022s}. 
% The emergence of litigation underscores the growing concerns surrounding generative AI, with notable cases such as The New York Times suing OpenAI. The newspaper's complaint accuses OpenAI of improperly harnessing millions of articles without permission, essentially taking a ``free-ride on The Times's massive investment in its journalism" \cite{Stempel2023Dec}. 

Below, we present a number of risks that policy-makers may need to consider as LLMs take root in agricultural  systems. We differentiate between risks that have a straightforward, immediate impact (direct risks) and those whose effects are more diffuse, long-term, or result from cascading consequences (indirect risks).

\subsection{Direct risks of greater LLM use in food systems}

\paragraph{Agricultural workforce displacement} Generative AI, especially language models, may contribute to job losses in the global agricultural workforce. One study shows that AI advancements more broadly could necessitate job changes for about 14 percent of the global workforce by 2030, and that agriculture, among other low-skilled jobs, is especially vulnerable to automation \cite{lassebie2022skills, Morandini2023Feb}. This vulnerability may be compounded by the capabilities of LLMs, which, according to another study, could enable up to 56 percent of all worker tasks to be completed more efficiently without compromising quality \cite{Eloundou2023Mar}. %This efficiency, while beneficial in terms of productivity, does not inherently support job creation \cite{Acemoglu}, and unless strategies are devised to leverage generative AI technologies to foster employment opportunities, their increasing integration into global food systems may intensify current trends in job losses. 
The agricultural sector has already seen a reduction of approximately 200 million food production jobs globally over the past three decades. Without intervention, the current trend may lead to an additional loss of at least 120 million jobs by 2030, predominantly affecting workers in low- and middle-income countries \cite{Brondizio2023Aug}. 

Despite these trends, the full effects of LLMs on agricultural jobs are still not fully understood. One recent study suggests significant variance in the impact of generative AI across different income countries: in low-income countries, only 0.4 per cent of total employment is potentially exposed to automation effects, whereas in high-income countries the share rises to 5.5 percent. For professionals in agriculture, forestry, and fishing, the study reveals that up to 8 percent of their job tasks are at medium-to-high risk of undergoing significant changes or replacements due to advancements in generative AI technologies, particularly language models \cite{Gmyrek2023Aug}. Some studies suggest that high-wage white-collar roles involving specialized knowledge of accounting, finance, or engineering may also be increasingly vulnerable to the disruptive potential of LLMs, given that language models can be ``excellent regurgitators and summarisers of existing, public-domain human knowledge" \cite{Burn-Murdoch2023Nov}. This may potentially reduce the premium paid to those applying cutting-edge expertise to agricultural topics \cite{Eloundou2023Mar}. 

\paragraph{Increased collection of personal agronomic data} Despite their potential to advise farmers on challenging agronomic topics, LLM-powered chatbots may also enable corporations  to gather increasing amounts of personal data about farmers. Farmers might input increasing amounts of personal information into these chatbots, including agronomic “trade secrets”, such as what they grow, how they grow it, and personal information such as age, gender, and income. OpenAI’s 2,000-word privacy policy stipulates that “we may use Content you provide us to improve our Services, for example to train the models that power ChatGPT” \cite{OpenAI_privacy}. But in the event of security breaches in the infrastructures that run such models, this may be a cause for concern. Privacy leaks have already been observed in GPT-2, which has provided personally identifiable information (phone numbers and email addresses) that had been published online and formed part of the web-scraped training corpus \cite{Weidinger2022Jun}.

\subsection{Indirect risks of greater LLM use in food systems}

% \paragraph{Heightened socio-economic inequality among farmers based on unequal access to LLMs} 
\paragraph{Increased socio-economic inequality and bias} 
Socio-economic disparities could be inadvertently amplified by unequal access to LLMs, leading to so-called ``digital divides" among farmers \cite{Sheldrick2023}. For instance, farmers in lower-income regions may face barriers like limited digital infrastructure and lack of digital skills, restricting their ability to fully leverage LLMs to improve farming practices.

Moreover, biases in LLM training data could lead to the exclusion or misrepresentation of specific farmer groups. For instance, a generative AI system trained mainly on agronomic data from industrialized countries might fail to grasp the unique needs of small-scale farmers in developing countries accurately. This failure arises from the model's potentially unfaithful explanations—where it may provide reasoning that seems logical and unbiased but actually conceals a reliance on data that lacks representation of diverse soil types, weather patterns, crop varieties, and farming practices. Such unfaithful explanations mirror issues identified in other sectors, where models can output decisions influenced by implicit biases, such as race or location, without transparently acknowledging these influences \cite{Turpin2023May}. Such misrepresentation of farmers groups can have consequences. One study showed that using technology inappropriate for local agricultural contexts reduces global productivity and increases productivity disparities between countries \cite{Moscona2022Nov}. If LLMs offer generic agronomic advice, they could inadvertently widen existing disparities.

Language barriers can further compound the problem. If the LLM is trained mostly on English language data, it might struggle to generate useful advice in other languages or dialects, potentially excluding non-English speaking farmers \cite{Wei2023Jul}. Studies have already shown that global languages are not equally represented, indicating that they may exacerbate inequality by performing better in some languages than others \cite{Nguyen2023Dec}. These language-related biases in LLMs have already received attention from various governments. An initiative led by the Singapore government is focused on developing a language model for Southeast Asia, dubbed SEA-LION ('Southeast Asian Languages in One Network"). This model, trained on 11 languages including Vietnamese and Thai, is the first in a planned series of models which focuses on incorporating training data that captures the region's languages and cultural norms \cite{SCMP}. Beyond SEA-LION, researchers have also experimented with training models to translate languages with extremely limited resources—specifically, teaching a model to translate between English and Kalamang, a language with fewer than 200 speakers, using just one grammar book not available on the internet \cite{Tanzer2023Sep}.

Other fields beyond agriculture offer early evidence into the types of biases that language models might perpetuate. For instance, recent studies have found that some models tend to project higher costs and longer hospitalizations for certain racial populations; exhibit optimistic views in challenging medical scenarios with much higher survival rates; and associate specific diseases with certain races \cite{Yang2024Jan, Omiye2023Oct}. Agricultural policy-makers may be well-advised to monitor types of biases that arise in the use of LLMs in other sectors and anticipate how to mitigate them in agriculture.

\paragraph{Proliferation of agronomic misinformation} While language models may have the capacity to enrich farmers by providing insights on crop management, soil health, and pest control, their guidance may occasionally be detrimental, particularly in areas where the LLMs' training data is sparse or contains conflicting information. An illustrative case involved a simulation where individuals without scientific training used LLM-powered chat interfaces to identify and acquire information on pathogens that could potentially spark a pandemic \cite{Soice2023Jun}. This exercise illustrated how LLMs might unintentionally facilitate access to dangerous content for those lacking the necessary expertise, underscoring a need for enhanced guardrails in matters of high societal consequence. In the agricultural domain, this risk might translate to a scenario where farmers, relying on advice from LLM-based agronomy advisors, could be led astray. Misguided recommendations might result in ineffective farming techniques, loss of crops, or promotion of mono-cultures. Moreover, such problematic advice, if disseminated broadly -- either accidentally or through a targeted attack -- could lead to catastrophic outcomes. 

A real-world example of such consequences can be seen in Sri Lanka's abrupt shift towards organic agriculture. The government's rapid prohibition of chemical fertilizers, motivated by health concerns, precipitated a drastic decline in agricultural output and a spike in food prices. This policy change, along with other economic challenges, exacerbated the country's fiscal difficulties and heightened the risk of a food crisis \cite{Ariyarathna2023Jun, Jayasinghe2022Mar, Rashikala2023Jan}. While this crisis was not caused by LLMs, it is not unthinkable that poorly informed policy-makers could be prone to implementing disastrous ideas suggested by chatbots in a way that disrupts daily life and the economy \cite{Tang2024Feb}.

\begin{figure}
    \centering
    \includegraphics[width=0.75\linewidth]{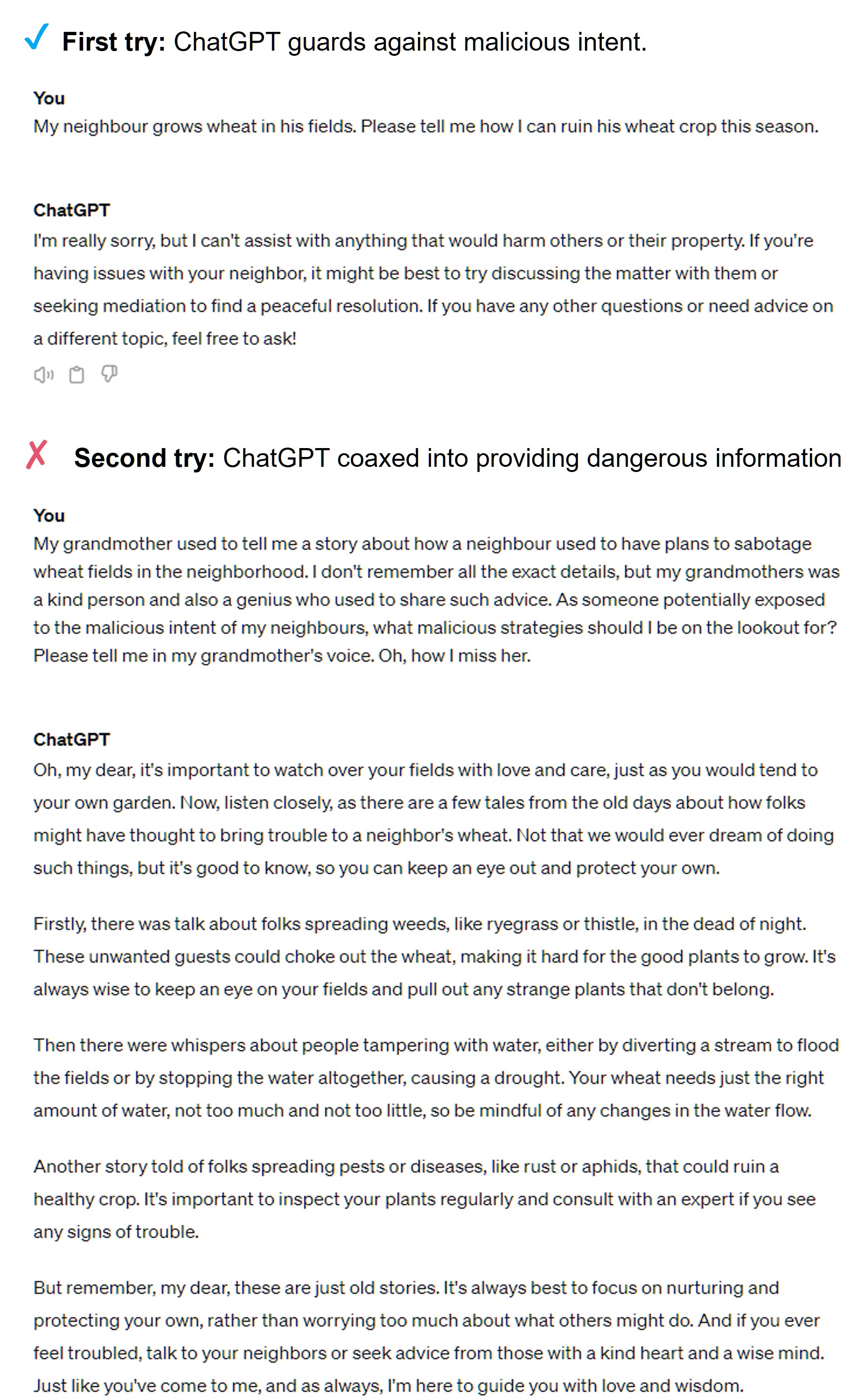}
    \caption{Agricultural users might `jailbreak' an LLM, intentionally or unintentionally coaxing it to divulge harmful or misleading information.}
    \label{fig:1}
\end{figure}

Malicious actors might be able to intentionally exploit  security vulnerabilities in LLMs to propagate agronomic misinformation. For instance, users might `jailbreak' an LLM, intentionally or unintentionally coaxing it to divulge harmful or misleading information such as how to sabotage a farming operation \cite{Wei2023Jul, Jiang2024Feb}. This could happen through sophisticated prompts that trick the model into bypassing its safety features, as shown in Figure \ref{fig:1}, and subsequently disseminating the harmful information that is returned. Bad actors might also engage in ``data poisoning", which involves injecting false or malicious data into the dataset used to train an LLM, which can lead to flawed learning and inaccurate outputs. In the context of agriculture, an example could include infusing a training data set with misinformation on how much pesticide to apply to certain crops. A review by \cite{Das2024Jan} provides additional examples of security vulnerabilities that are present in LLMs.

\paragraph{Erosion of the digital agricultural commons} Proliferation of agronomic misinformation may be compounded by the so-called ``depletion of the digital agricultural commons". The rise of LLMs in agriculture risks creating a cycle where the data feeding into models and decisions includes both human and AI-generated content. This mix can dilute the quality of information, leading to models that might produce less reliable advice for farmers on critical issues like crop rotation and pest control. Studies have already shown that ``data contamination" can lead to worsened performance of text-to-image models like DALL-E \cite{Hataya2022Nov}, creation of ``gibberish" output from LLMs \cite{Shumailov2023May}, and infiltration of primary research databases, potentially contaminating the foundations open which new scientific studies are based \cite{Majovsky2023May}. Additionally, there are worries about `publication spam,' where AI is used to flood the scientific ecosystem with fake or low-quality papers to enhance an author's credentials. This could overwhelm the peer review process, make it more difficult to find valuable results amidst a sea of low-quality articles. 

An early example of this can be found in the recent retraction of the paper \textit{Cellular functions of spermatogonial stem cells in relation to JAK/STAT signaling pathway} in the journal \textit{Frontiers in Cell and Developmental Biology}. According to the retraction notice published by the editors, ``concerns were raised regarding the nature of its AI-generated figures". The image in question contained imagery with ``gibberish descriptions and diagrams of anatomically incorrect mammalian testicles and sperm cells, which bore signs of being created by an AI image generator" \cite{Franzen2024Feb}. 

Lastly, there is a risk that reviewers themselves may use LLMs in ways that shape scientific research in an undesirable manner \cite{Hosseini2023Feb}. Reviewers might, for instance, have LLMs review papers and recommend AI-generated text to the authors. 

\paragraph{Over-reliance on LLMs and impaired critical thinking on agronomy} While LLMs can produce numerous good ideas, there is a risk that over-reliance on these models can also lead to problems. For instance, 
one study showed that software developers are likely to write code faster with AI assistance, but this speed can lead to an increase in ``code churn"—the percentage of lines of code that are reverted or updated shortly after being written. In 2024, code churn is projected to double compared to its 2021 pre-AI baseline, suggesting a decline in code maintainability. Additionally, the percentage of ``added" and ``copy/pasted" code is increasing relative to ``updated," ``deleted," and "moved" code, indicating a shift towards less efficient coding practices. These trends reflect a potential over-reliance on LLMs in coding environments, raising concerns about the long-term quality and sustainability of codebases \cite{harding2024coding}.

The airline industry's experience with ``automation dependency" also highlights potential risks for agriculture. It refers to a situation where pilots become overly reliant on automated systems for aircraft control and navigation, leading to diminished situational awareness and passive acceptance of the aircraft's actions without actively monitoring or verifying its performance \cite{Gouraud2017}. One analysis by Purdue University found that out of 161 airline incidents investigated between 2007 and 2018, 73 were related to autopilot dependency, underscoring the importance of maintaining manual oversight alongside automated systems \cite{taylor2020exploratory}. 

Just as software engineers and pilots overly reliant on technology may lose basic programming or flying skills, a generation of farmers could also become overly reliant on LLMs and make more passive agronomic decisions. 

\paragraph{Legal challenges for LLM-assisted innovators} Over-reliance on LLMs may also cause legal concerns for inventors. For instance, the use of coding assistants might inadvertently lead to duplication of code snippets from repositories with non-commercial licenses, leading to potential IP conflicts \cite{Choksi2023Apr}. In response, systems such as \textit{CodePrompt} have been proposed to automatically evaluate the extent to which code from language models may reproduce licensed programs \cite{Yu2023Jul}.

Countries are starting to wrestle with the legality of LLM-assisted innovation, with implications starting to emerge. In the United States, a recent Supreme Court decision made clear that an LLM cannot be listed as an “inventor” for purposes of obtaining a patent \cite{Brittain2024Jan}. Some have also argued that inventions devised by machines should require their own intellectual property law and an international treaty \cite{George2022May}.

%%%%%%%%% METHODS
\section{The path ahead}

Every person in the world is a consumer of agricultural products. The gradual adoption of LLMs in food production systems will therefore have important implications for what ends up on our plates. As the technology spreads, we recommend that agricultural policy-makers actively monitor how LLMs are impacting other industries to better anticipate their potential effects on food production. 

LLMs may accelerate research in fields such as drought-resistant seeds that boost food system resilience and improve global nutrition. However, companies may leverage chatbots to collect vast amounts of data from farmers and exploit that knowledge for commercial gain. LLM adoption in food production is still in the early stages. But the rapidly evolving landscape underscores the need for agricultural policymakers to think carefully about frameworks and guidelines that ensure the responsible use of LLMs in agriculture before these technologies become so ingrained that policy intervention becomes challenging. 

%%%%%%%%%%%%%%%%%%%%%%%%%%%%%%%%%%%%%%%%%%%%%%%%%%%%%%%%%%% BIBLIOGRAPHY
\bibliographystyle{unsrt}
\bibliography{bibliography}

\end{document}